\documentclass[11pt,a4paper]{article}

\usepackage[T1]{fontenc}
\usepackage[utf8]{inputenc}
\usepackage{lmodern}
\usepackage{microtype}
\usepackage[margin=25mm]{geometry}
\usepackage{graphicx}
\usepackage{booktabs}
\usepackage{tabularx}
\usepackage{array}
\usepackage{enumitem}
\usepackage[font=small,labelfont=bf]{caption}
\usepackage[numbers,sort&compress]{natbib}
\usepackage{xcolor}
\usepackage{pdflscape}
\usepackage[colorlinks=true,linkcolor=blue!50!black,citecolor=blue!50!black,urlcolor=blue!50!black]{hyperref}

\newcolumntype{L}{>{\raggedright\arraybackslash}X}
\newcommand{\atlasurl}{\url{https://vitaldb.metrology-atlas.org}}

\title{The Metrology Atlas: describing recorded physiological data as a measurement system, with VitalDB as a first application}
\author{Andrew J. Goodwin\\
  \small School of Biomedical Engineering, The University of Sydney}
\date{October 2026}

\begin{document}
\maketitle

\begin{abstract}
Every value in a recorded physiological dataset is a pair: a recorded time and a recorded value. Neither is the
truth. The time comes from a clock that drifts and is corrected in jumps, and often describes a window that ended
earlier. The value was quantised, rounded, sometimes heaped by hand, and measured by an instrument with its own
error. We present the Metrology Atlas, a dataset-independent way of describing a recorded dataset as a measurement
system, and its first application, to VitalDB, an open dataset of 6,388 surgical patients, as a proof of principle.
For every variable, the atlas states what was measured, the measurement model linking it to the recorded pair, and what is known about
the true time and true value given a record. Each part of the model carries its evidence: 69 studies, 49
parameters and 54 observations, each marked as measured, derived, estimated or assumed. The atlas maps 11 clocks
and 14 devices into one timing network, and closes loops in that network to test its own model. Synchronisation
reduces the median timing error between the EEG and ECG from 794\,ms to 42\,ms, and between ventilator and monitor
waveforms from 378\,ms to 55\,ms. The atlas also documents data-quality issues, several of which correct the
dataset's own documentation. It is published as a website and a single machine-readable file, with open-source
code and a scheme of time-aware labels. Everything is open under CC~BY~4.0 and MIT licences. The engine, schema
and timing tools carry over unchanged, and the aim is to describe many other datasets the same way.
\end{abstract}

\section{Motivation}

\begin{figure}[t]
  \centering
  \includegraphics[width=\textwidth]{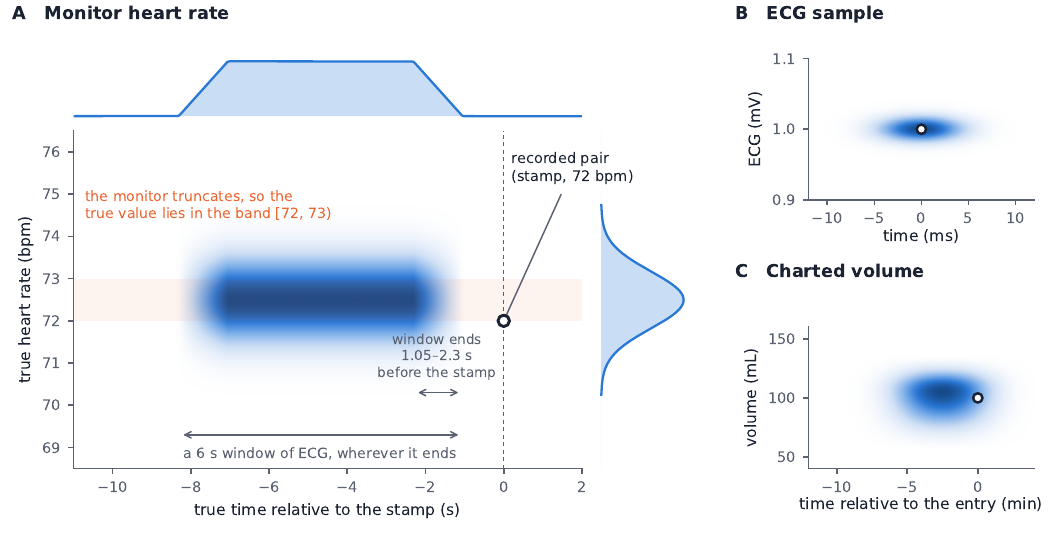}
  \caption{A record is a point; what it tells us is a region. Each panel shows what is known about the true time
    (horizontal) and true value (vertical) behind one recorded pair (white dot), with darker shading where the truth
    is more likely; the curves in A are its two marginals. (A) A monitor heart rate of 72\,bpm describes a 6\,s window
    of ECG that ended 1.05--2.3\,s before its stamp, and the monitor truncates, so the true value lies in
    [72,\,73) blurred by measurement error. (B) An ECG sample is known to milliseconds and microvolts. (C) A charted
    volume entered on a 5-minute mark is uncertain by minutes, and heaping makes its true value broad and skewed.
    Shapes follow the VitalDB atlas's models; widths are illustrative.}
  \label{fig:concept}
\end{figure}

Open physiological datasets are now the training ground for clinical machine learning and time-series foundation
models. Their users mostly treat each record as exact: a value that happened at its time stamp. In a multi-device
recording such as VitalDB, that assumption fails in ways that matter. Device clocks drift by tens to hundreds of parts per million
against the recording PC and are periodically reset to it, so their time error grows and then jumps back: a
sawtooth. Monitor numerics describe a window of
signal that ended seconds before their stamp. Hand-charted events are rounded to five-minute marks. Converters
store fewer bits than documented. None of this is visible in a typical loader's output, and an analysis that
relates two signals in time (pulse arrival time, EEG--haemodynamic coupling, an alarm against its cause) inherits
all of it (Figure~\ref{fig:concept}).

Timing errors and temporal uncertainty in clinical databases have been reviewed before~\citep{goodwin2022timing},
and specific cases have been solved, such as aligning EEG with bedside waveforms~\citep{goodwin2023}. What is
missing is a complete, checkable description of one dataset's measurement system that others can read, reuse and
correct. Metrology already has the language for this: a measurand, a measurement model, and a stated uncertainty
with its evidence~\citep{jcgm-200-2012,jcgm-100-2008}. The atlas applies that language to recorded data. The
measurements were made long ago, so every part of the model is reconstructed from the records themselves.

The Metrology Atlas is a general approach, not a VitalDB project. Its schema, engine and timing tools know
nothing about any one dataset; a dataset is described by a folder of its own. VitalDB is the first, chosen as a
proof of principle because it is open, large, and records many devices on separate clocks. The goal is a family of
atlases, one per dataset, sharing one schema, so that the measurement properties of different datasets can be
read, compared and reused in the same way.

This paper introduces the approach through the VitalDB atlas: the themes it investigates, what it found, and how
to use and extend it. The
detail lives in the atlas itself; this paper is a guide to it. While the approach is general, the numbers in
this paper, including the examples above, are measured in VitalDB, on its devices and recording set-up, unless
stated otherwise. They show the kinds of error the approach describes, not values to assume for another
dataset, which needs its own measurements. The values and results quoted here are correct at
the time of publication and match the archived snapshot of the atlas cited in Section~\ref{sec:access}. The atlas
itself is a living document: studies are added, models refitted and numbers revised as evidence comes in, so the
website may have moved on since. Superseded results stay on the site, marked as such, so any number quoted here can still be traced.

This preprint accompanies an open, pre-release version of the atlas, published for review by the community and
by the dataset's custodians ahead of a peer-reviewed journal version. Comments, corrections and new evidence are
invited (Section~\ref{sec:contribute}).

\section{Intended audience}

The atlas is written for four groups:
\begin{itemize}[itemsep=2pt]
  \item \textbf{Researchers using VitalDB}, who need to know how far apart two signals' clocks can be, what a
    numeric's delay is, or whether a charted time was rounded.
  \item \textbf{Builders of models and benchmarks}, who need time-aware labels that respect when a value could
    first have been known, and need to know which distributions are artefacts of rounding or heaping.
  \item \textbf{Dataset curators and device makers}, for whom the VitalDB atlas is a worked example of documenting a
    recording system and a list of specific corrections, and the engine a way to describe their own datasets.
  \item \textbf{Metrologists and data-quality researchers}, who may find the framing useful for other datasets;
    the engine is dataset-agnostic.
\end{itemize}

\section{The VitalDB dataset}

VitalDB holds high-resolution waveforms, numerics and clinical information from 6,388 non-cardiac surgical
patients at Seoul National University Hospital~\citep{lee2022vitaldb}. The atlas describes version 1.0.0 on
PhysioNet~\citep{lee2022vitaldb-physionet,pollard2026physionet}, released under CC~BY~4.0 with open access and
no data use agreement.

Each operating room had one PC running Vital Recorder~\citep{lee2018vitalrecorder}, which stamped every record from
every device on arrival. The waveforms that matter most for timing come through SNUADC, a 500\,Hz converter that
samples the patient monitor's analogue outputs (ECG, arterial pressure, plethysmogram, CVP). The BIS monitor, the
Primus anaesthesia machine and several haemodynamic monitors send their own waveforms and numerics, each on its
own clock. Clinical events and laboratory results come from the hospital's records, on yet another clock. The
release shifts each case to a common start and shuffles case ids, so absolute dates and times of day are not
available.

\section{The Metrology Atlas}

The atlas is a folder of YAML files, one per item, from which an engine builds a website and a single
machine-readable file (\texttt{vitaldb.metrology.json}). The engine knows nothing about VitalDB; another dataset is another folder.

What carries over to a new dataset is the schema (the kinds of item and how each number states its evidence), the
engine and its checks, the website and machine-readable output, the synchronisation planner, and the
dataset-independent timing tools in physiosync (stamp clocks, corrected times, the clock-network adjustment). What
is written afresh is the dataset's own folder: its variables, clocks and devices, and the studies that measure them,
with the analysis code behind them. Each dataset's atlas has its own address under the same domain, as
\texttt{vitaldb.metrology-atlas.org} does for VitalDB.

Items are of eleven kinds. \textbf{Variables} are the recorded tracks and fields. \textbf{Clocks}, \textbf{links}
and \textbf{devices} describe how time is carried, from UTC through the recorder PC to each device.
\textbf{Parameters} are the numbers the models rely on, each with a status (measured, derived, estimated, assumed,
withdrawn or superseded) and the study that produced it. \textbf{Studies} are short reports: background, data in
aggregate, method, results, limitations and the code that ran them. They are grouped by \textbf{method} and built
from shared \textbf{techniques}. \textbf{Observations} are general facts about the data, each established, likely,
open or withdrawn. \textbf{Software} and \textbf{references} complete it. Each fact lives in one place, and every
relation is written once and linked both ways.

\begin{figure}[t]
  \centering
  \includegraphics[width=0.9\textwidth]{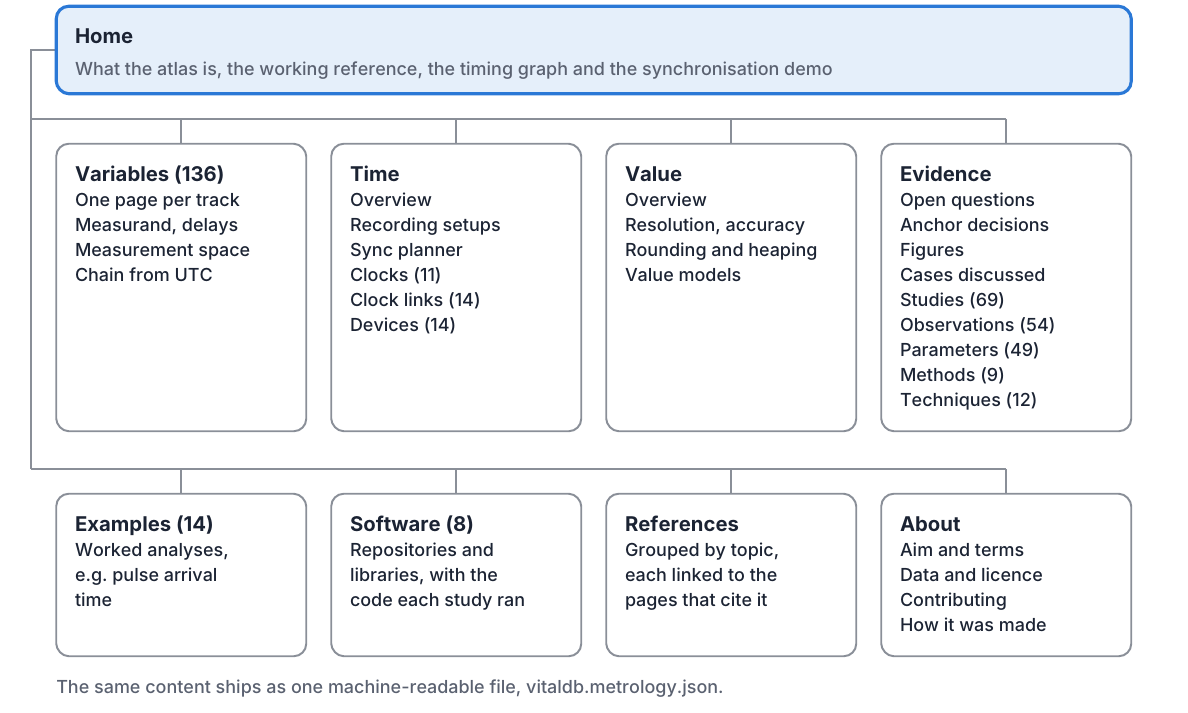}
  \caption{The atlas website. Every item is one page, linked both ways to the items it relates to; counts are for
    the current release.}
  \label{fig:sitemap}
\end{figure}

For every variable, the site draws a measurement-space figure (where the true time and true value can lie, given a
record) and a chain from UTC to the stored time stamp. A synchronisation planner combines any two variables'
timing to give the expected misalignment between them. Open questions are collected from every study and classed
as phenomenon, hypothesis, assumption, conflict or next step. The current release holds 136 variables, 11 clocks,
14 links, 14 devices, 49 hand-written parameters, 69 studies, 54 observations, 9 methods and 12 techniques
(Figure~\ref{fig:sitemap}).

\section{Themes of investigation}

The atlas treats each record as an indication of two measurands, when something happened and what its value was,
and asks the same questions of both axes. Each axis has two kinds of error: accuracy (how far the record sits from
the truth, systematically) and resolution (how finely it can be read at all). The studies fall into nine themes
(Table~\ref{tab:themes}).

\begin{table}[t]
  \centering
  \small
  \caption{Themes of investigation, each with the question it asks and an example finding.}
  \label{tab:themes}
  \begin{tabularx}{\textwidth}{@{}p{0.2\textwidth}LL@{}}
    \toprule
    Theme & Question & Example finding \\
    \midrule
    Quantisation and resolution & How finely was each value and time recorded? &
      SNUADC stores 10-bit codes, not the documented 12; a time on a grid of step $\Delta$ carries
      $\Delta/\sqrt{12}$ of error \\
    Rounding and heaping & Was a value rounded, truncated or heaped by hand? &
      Urine output is heaped: 43\% of values sit on 50 or 100\,mL, so a recorded 100\,mL is uncertain by about
      $\pm$20--40\,mL; anaesthesia start and end entered on 5-minute marks were rounded up \\
    Clocks and sensors & How fast does each clock drift, and how is it corrected? &
      SNUADC $-63$\,ppm, BIS $+25$\,ppm, Primus $+104$\,ppm, all against the recorder PC \\
    The recording system & Which clock stamps which record, through which path? &
      11 clocks in one timing network; drift is largely set by the setup (the PC and its ports) \\
    Delays and windows & What interval does a stamped value describe? &
      Every numeric describes a window ending 1--9\,s before its stamp; the BIS index averages 15--60\,s of EEG \\
    Synchronisation & How well can two devices' times be brought together? &
      EEG to ECG: 42\,ms median after synchronisation, from 794\,ms \\
    Continuity & Where is signal missing, duplicated or reordered? &
      21\,s of airway pressure lost every 7,230\,s; five records reordered at every SNUADC reset \\
    Availability & When could a value first have been known? &
      Laboratory time stamps mark the specimen, not the result \\
    Invertibility & Given a record, what can be said about the truth? &
      Drift is invertible once measured; rounding and truncation are not, and leave a distribution \\
    \bottomrule
  \end{tabularx}
\end{table}

\begin{figure}[t]
  \centering
  \includegraphics[width=\textwidth]{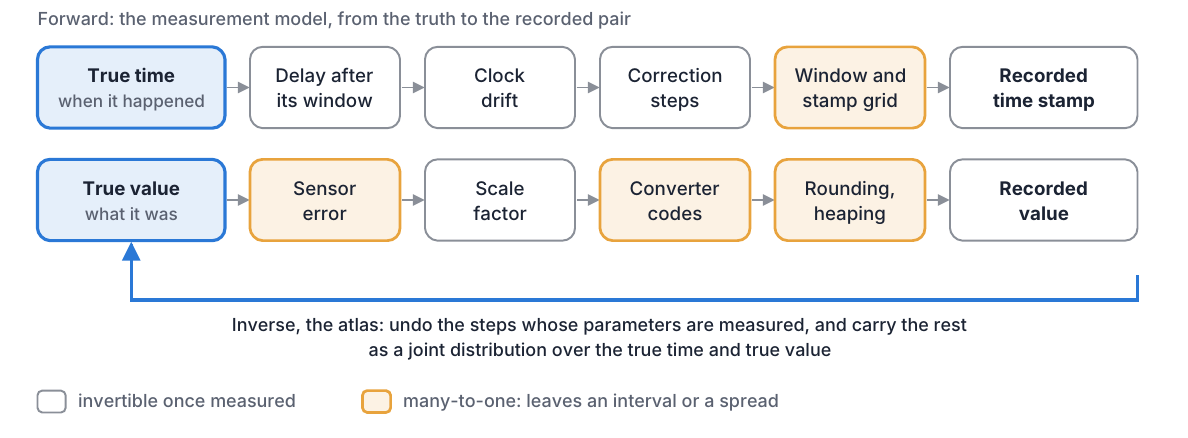}
  \caption{The measurement model runs forward from the true time and value to the recorded pair. Steps whose
    parameters can be measured (delay, drift, correction steps, scale factors) can be undone; steps that are
    many-to-one (a window and a coarse stamp grid, sensor error, converter codes, rounding and heaping) cannot, so
    inverting a record gives a distribution rather than a point.}
  \label{fig:invertibility}
\end{figure}

\begin{landscape}
\begin{figure}
  \centering
  \includegraphics[width=\linewidth,height=0.84\textheight,keepaspectratio]{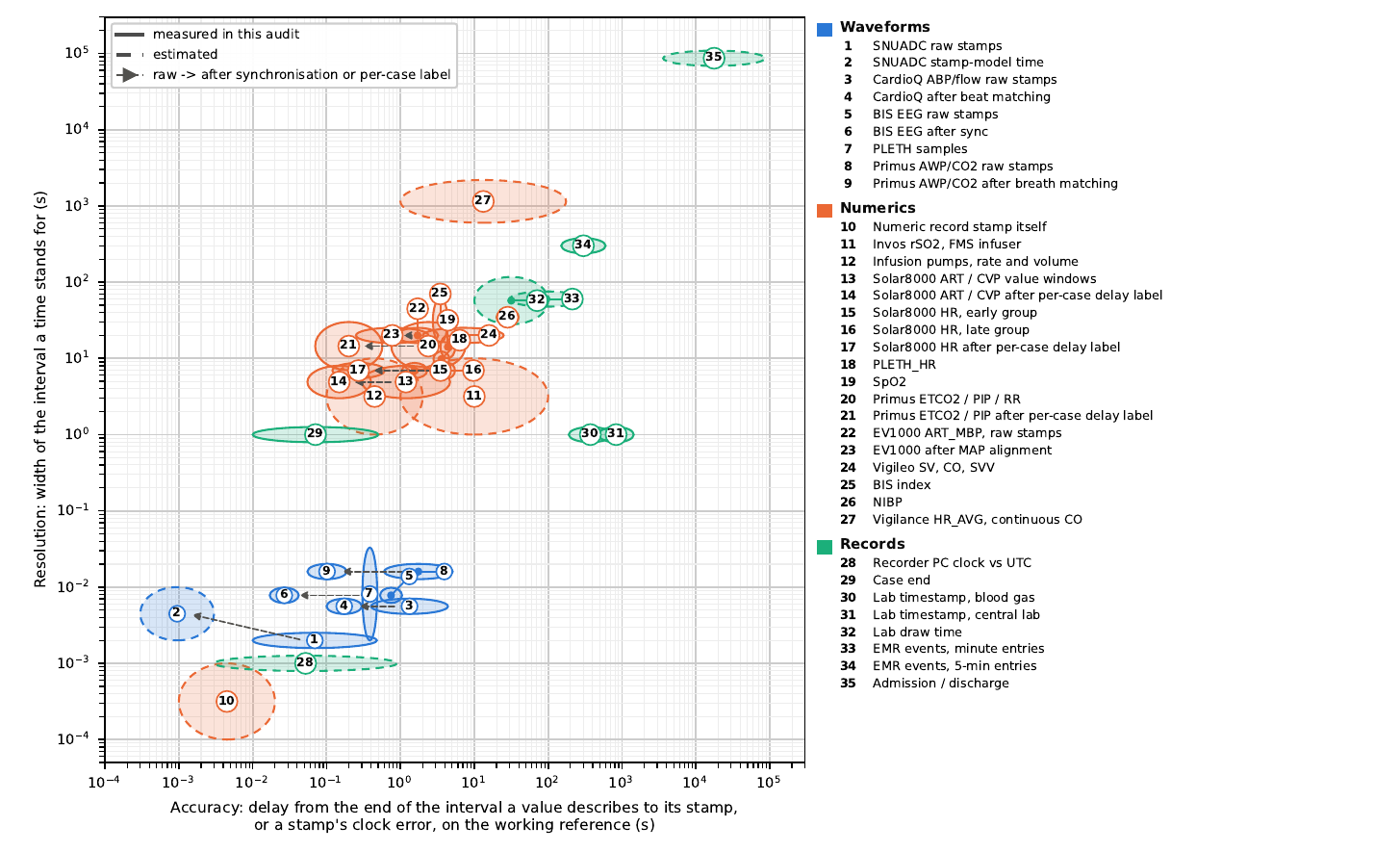}
  \caption{Accuracy against resolution for every kind of time stamp in VitalDB, on log axes. Each ellipse spans
    the measured (solid) or estimated (dashed) range; arrows join a raw stamp to the same stamp after
    synchronisation or a per-case label. Waveform samples (blue) sit at millisecond resolution; numerics (orange)
    describe windows of seconds that ended seconds earlier; records and hand-entered times (green) span minutes to
    days.}
  \label{fig:accuracy}
\end{figure}
\end{landscape}

\textbf{Invertibility} ties the themes together (Figure~\ref{fig:invertibility}). A measurement model runs forward,
from the true time and value to the recorded pair, through steps such as delay, drift, correction, quantisation
and rounding. The atlas runs it backwards. Some steps are invertible once their parameters are known: a measured
drift can be removed. Others are many-to-one: a rounded value or a stamp on a coarse grid maps back to an
interval, not a point. The result of inverting a record is therefore a joint distribution over its true time and
true value, in the sense of the GUM's Supplement~2~\citep{jcgm-102-2011}, which can be propagated through any later
calculation~\citep{jcgm-101-2008}. Figure~\ref{fig:accuracy} places every kind of time stamp in VitalDB on the two
axes of accuracy and resolution.

\section{Signal synchronisation}

\begin{figure}[t]
  \centering
  \includegraphics[width=\textwidth]{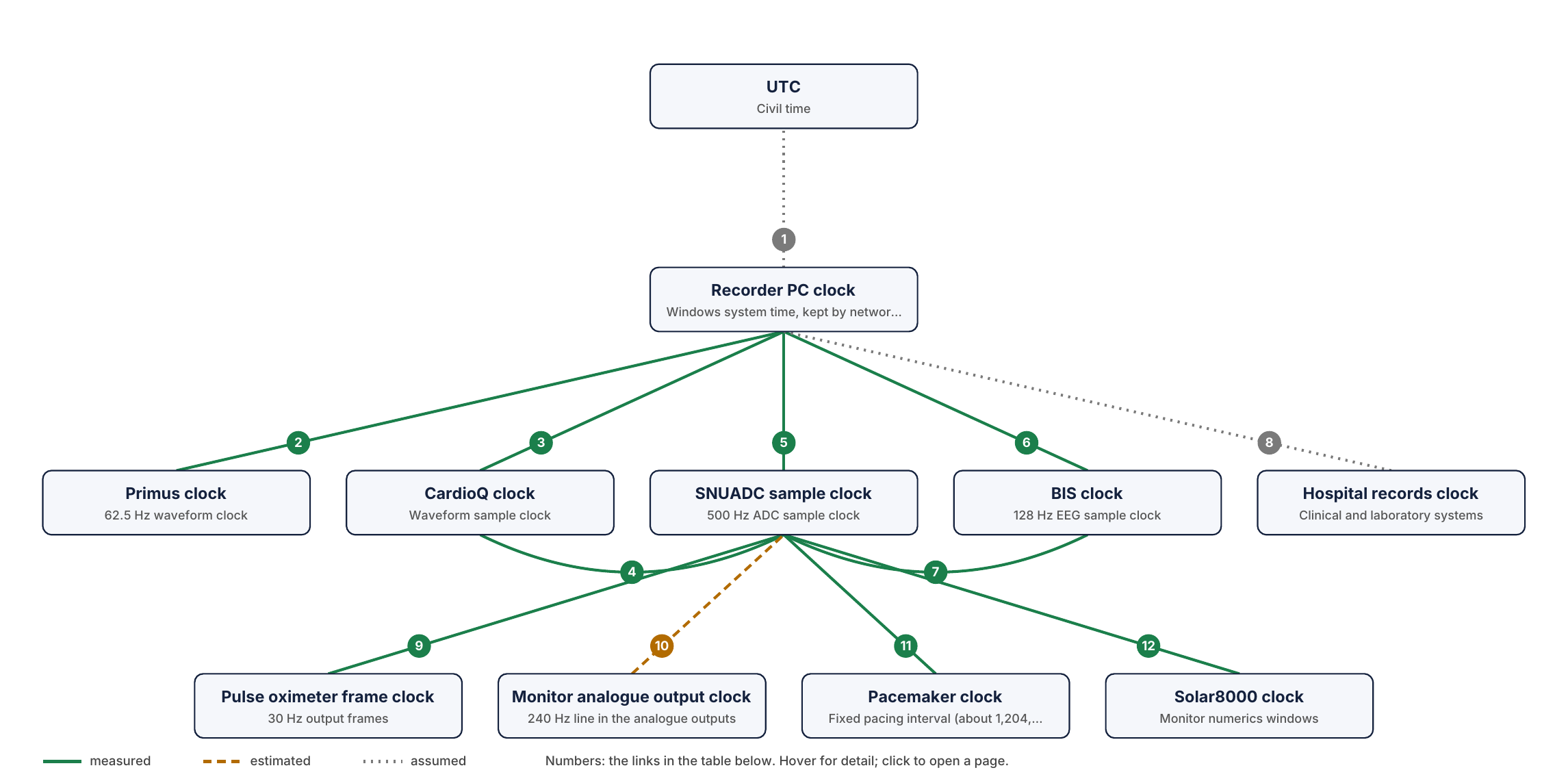}
  \caption{The VitalDB timing network. Each clock sits below the clock it is measured against; solid links are measured,
    dashed estimated, dotted assumed. The recorder PC's link to UTC (1) and the hospital records' link to the PC
    (8) cannot be observed in the release. Generated from the atlas on every build.}
  \label{fig:network}
\end{figure}

The VitalDB timing network places every clock one level below the clock it is measured against
(Figure~\ref{fig:network}). All device rates are relative: each is measured against another clock that also
drifts. The recorder PC's link to UTC cannot be observed in the release and is assumed, so no time in VitalDB is
traceable to UTC.

Three kinds of evidence carry the links. \textbf{Stamp reading} fits each device's drift and correction schedule
from its own time stamps. The resulting stamp model rebuilds each device's timeline on the recorder PC, including
the 17.7\,ms of signal lost at each SNUADC reset. \textbf{Shared events} are anything two devices both record,
whether it comes from the patient or from the equipment: electrocautery and muscle bursts in the ECG and the EEG,
the heartbeat's faint electrical trace in the EEG, or a breath in the airway pressure and the central venous
pressure. \textbf{Clocks carried in the signal} are regular rhythms inside a recorded signal that act as a third,
independent clock, such as the oximeter's frame slips in the plethysmogram. The atlas's working reference is SNUADC stamp-model
time, against which every delay is stated.

\begin{figure}[t]
  \centering
  \includegraphics[width=\textwidth]{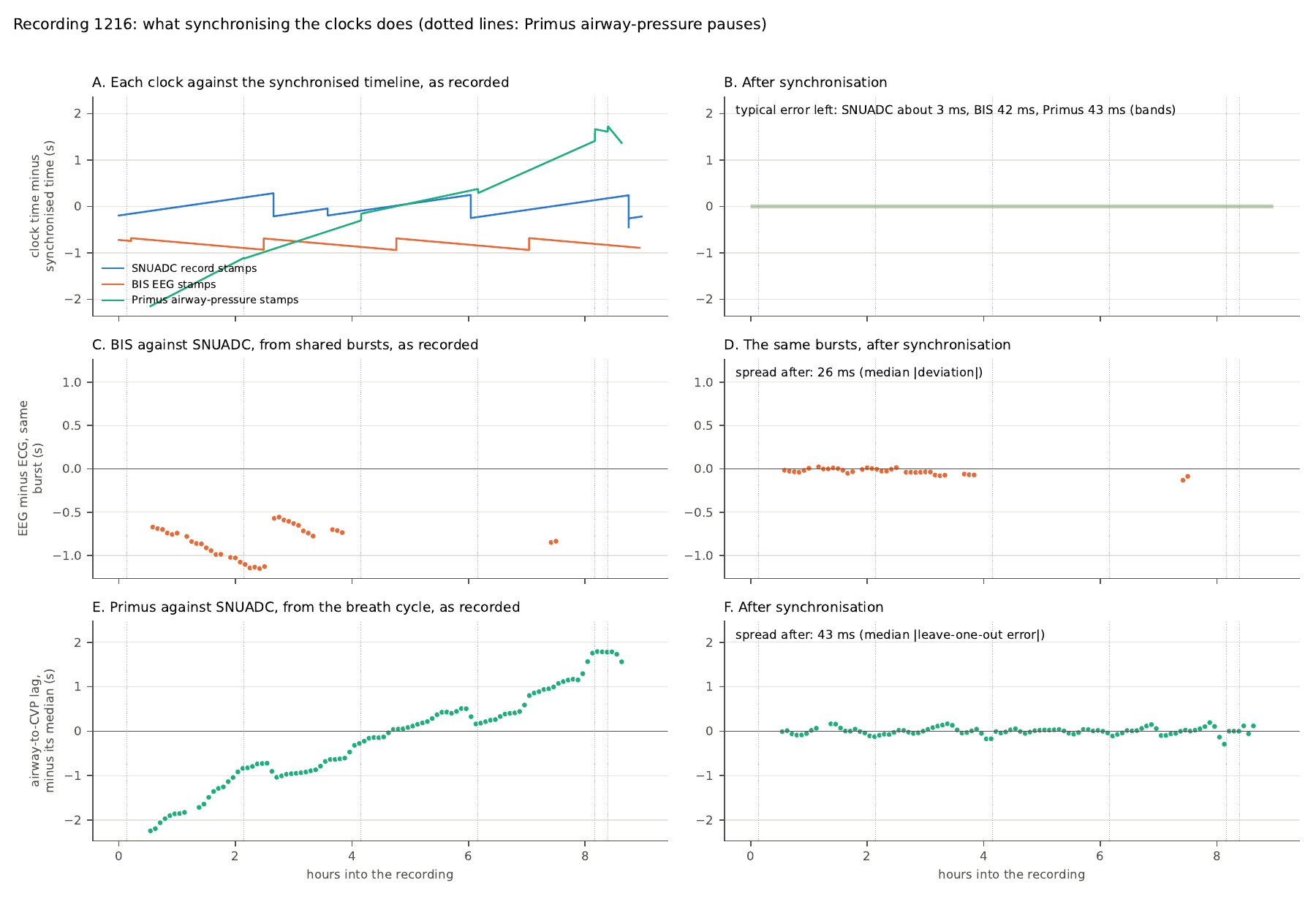}
  \caption{Residual timing errors between device pairs through one recording (case 1216, 8.7\,h), as recorded
    (left) and after synchronisation (right).}
  \label{fig:syncdemo}
\end{figure}

\begin{figure}[t]
  \centering
  \includegraphics[width=\textwidth]{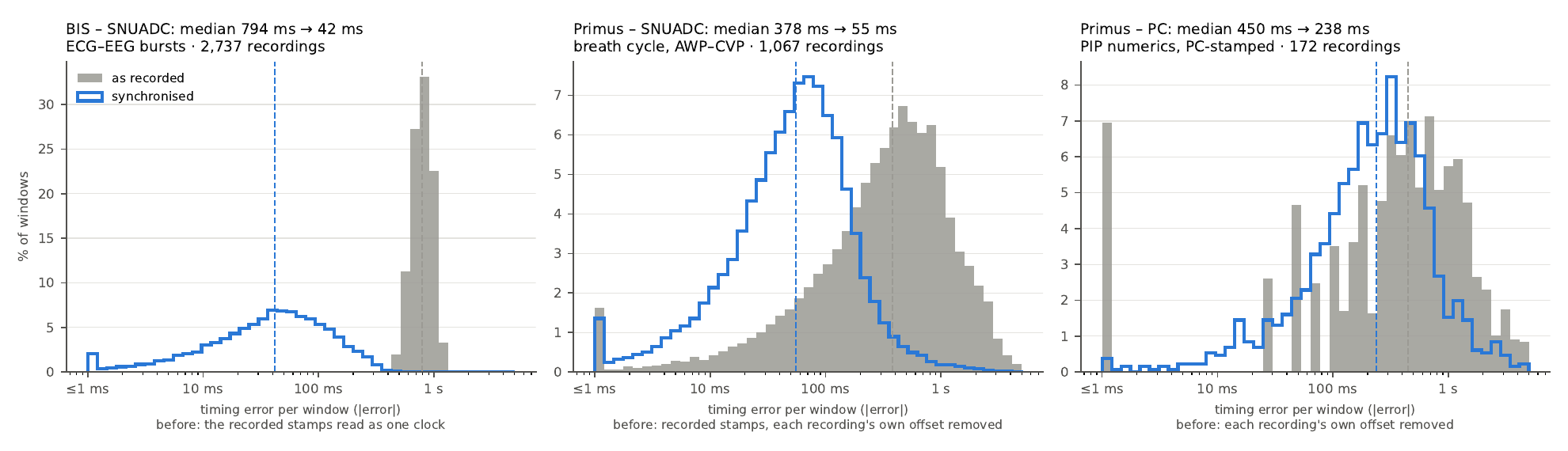}
  \caption{Timing error per 10-minute window across the release, as recorded (grey) and after synchronisation
    (blue), on a log scale; dashed lines mark the medians. BIS against SNUADC is checked by electrocautery and
    muscle bursts seen in both the ECG and the EEG (2,737 recordings); Primus against SNUADC by the breath cycle in
    airway pressure and CVP (1,067 recordings; each window predicted from the others with a line per stretch
    between airway-pressure pauses); Primus against the PC by PC-stamped PIP values (172 recordings,
    30-minute blocks). The checking signal never enters the correction it tests.}
  \label{fig:census}
\end{figure}

Figure~\ref{fig:syncdemo} shows what synchronisation does in one recording. Before correction, the EEG--ECG and
ventilator--monitor offsets wander by hundreds of milliseconds and jump at each device's corrections. After it,
they sit near zero. Across the release, the median error falls from 794\,ms to 42\,ms for the BIS EEG against the
ECG (2,737 recordings), and from 378\,ms to 55\,ms for the Primus airway pressure against the monitor
(Figure~\ref{fig:census}). The Primus figure is a leave-one-out accuracy over 1,067 recordings with CVP, with a
line fitted in each stretch between the ventilator's airway-pressure pauses; its 90th percentile is 187\,ms.

Where two routes connect the same pair of clocks, they form a loop, and the rates around a loop should sum to zero.
The atlas treats the network as a geodetic adjustment: each link is an edge with a value and a standard error;
fundamental cycles give misclosures; weighted least squares gives adjusted rates and normalised residuals that flag
a bad link. Over 253 loops through the BIS, the median misclosure is $+1.3$\,ppm; over 166 through the Primus,
$-2.9$\,ppm. Variance factors of 2 to 5 show that the stated standard errors are still too small, which is the
next thing to fix. The adjustment is implemented generically in physiosync, so it applies to any recording system
with a known clock graph. Further loops are planned through shared artefacts: ECG-derived respiration, breathing
and pacing spikes in the EEG, mains-frequency wander, cuff inflations and the Solar8000 numerics.

\section{Data-quality issues found}

Of the 54 observations, 39 are established, 7 likely, 7 open and 1 withdrawn. Several differ from VitalDB's
documentation (Table~\ref{tab:quality}); the atlas's import already corrects the scale factors. The VitalDB team knows this work is being released and will
review these findings once the atlas is public.

\begin{table}[t]
  \centering
  \small
  \caption{Issues that differ from VitalDB's documentation or from common assumptions.}
  \label{tab:quality}
  \begin{tabularx}{\textwidth}{@{}p{0.27\textwidth}p{0.24\textwidth}L@{}}
    \toprule
    Issue & Documented or assumed & Found \\
    \midrule
    SNUADC converter resolution & 12-bit & 10-bit codes (0--1023) in every channel \\
    Primus CO\textsubscript{2} gain & 0.1 & 0.075 \\
    Primus airway pressure scale & 0.1064\,/\,$-22$ and 0.01\,/\,0 & 0.1\,/\,$-20$ \\
    Primus airway pressure continuity & Continuous & About 21\,s lost every 7,230\,s, in 5,502 recordings \\
    Anaesthesia start & At induction & A median 14\,min before induction; 5-minute entries rounded up \\
    SNUADC reset records & In time order & Five records after each reset written out of order \\
    Laboratory time stamp & Result time & Specimen time; the result came later, by an unrecorded turnaround \\
    Operation end & Charted & Filled in 10\,min before the recording end in 537 cases (8.4\%) \\
    \bottomrule
  \end{tabularx}
\end{table}

Others concern the files rather than the documentation: recorder pauses in 29\% of files, byte-for-byte duplicated
records in 8 recordings, 12 recordings stored at 100\,Hz, three arterial tracks carrying the plethysmogram's scale
factors, and loaders that lose about 0.48\,s at every SNUADC reset when they place samples by stamp. None of this
is a criticism of VitalDB, which remains the most detailed public intraoperative dataset; most of these issues
would be found in any multi-device recording that was examined this closely.

\section{Software and labels}

\paragraph{Software.} The atlas is one of five repositories released together (Table~\ref{tab:repos}), and the
code is split by how far it generalises. Three know nothing about VitalDB. The metrology-atlas engine checks a
folder of YAML files against the schema and compiles it into the website and the machine-readable file; the
VitalDB atlas is one such folder in the same repository. physiosync holds the timing tools: it models a device's
stamp clock from its time stamps, returns corrected times, and solves the clock-network adjustment used to close
loops. It reads no file format and works on the times and samples it is given, so any dataset's reader can call
it. lineaccess finds accesses to invasive pressure lines (blood draws, flushes and zeroing) from the pressure
waveform alone. Two repositories are specific to VitalDB. vitaldb-bench holds the analyses behind most studies,
each a script that runs per recording on the analysis machine and publishes only aggregates, together with the
VitalDB clock model, corrected readers and the labels described below. vitaldb-resolution turns a recorded VitalDB
value or time into a distribution over its true value, using versioned model files; its engine is general and is
to be released separately (Section~\ref{sec:beyond}). Every study in the atlas links to the code that produced it,
and every value model to the version it shows, so a number can be traced to the script that made it.

\begin{table}[t]
  \centering
  \small
  \caption{Repositories released with the atlas.}
  \label{tab:repos}
  \begin{tabularx}{\textwidth}{@{}lLl@{}}
    \toprule
    Repository & What it holds & Licence \\
    \midrule
    metrology-atlas & The engine and the VitalDB atlas (YAML, figures, import scripts) & MIT (engine); CC~BY~4.0 (content) \\
    physiosync & Dataset-independent timing tools: stamp clocks, corrected times, the clock-network adjustment & MIT \\
    vitaldb-bench & The analyses behind most studies, the clock model, corrected readers and labels & MIT \\
    vitaldb-resolution & Value models: rounding modes, heaping, charted times, versioned and locked per release & MIT \\
    lineaccess & Detection of arterial-line accesses, zeroing and flushes & MIT \\
    \bottomrule
  \end{tabularx}
\end{table}

\paragraph{Labels.} Derived results are published as time-aware labels in 13 families, defined once in
vitaldb-bench. They cover plethysmogram timing (frame-clock sawtooth fits and offsets), arterial-line events
(accesses, zeroing, flushes, suspect pressure), laboratory draws and times, numeric delays, heart-rate delay groups
and BIS smoothing. Every label carries a span, its parameters and an \texttt{available\_at}: the earliest point its
information could have been known (the end of its span, the end of the track or case, a population constant, or a
laboratory time plus an unrecorded turnaround). This enforces an availability rule, that no value is used before
it could have been known, and prevents a model built on the labels from leaking information from the future.

\section{Access, licensing and use}
\label{sec:access}

The atlas, its code and an archived snapshot are listed in Table~\ref{tab:access}.

\begin{table}[t]
  \centering
  \small
  \caption{Where to find the atlas, its code and the data.}
  \label{tab:access}
  \begin{tabularx}{\textwidth}{@{}p{0.28\textwidth}L@{}}
    \toprule
    Resource & Location \\
    \midrule
    The atlas website & \atlasurl \\
    The machine-readable atlas & \texttt{vitaldb.metrology.json}, on the site and in each release \\
    Code repositories & \url{https://github.com/metrology-atlas} \\
    Archived release & Zenodo snapshot of the version this paper describes (YAML source, JSON file and built site),
      \href{https://doi.org/10.5281/zenodo.23494769}{doi:10.5281/zenodo.23494769}; the concept DOI \href{https://doi.org/10.5281/zenodo.23494768}{doi:10.5281/zenodo.23494768} resolves to the
      latest version \\
    VitalDB 1.0.0 & \url{https://physionet.org/content/vitaldb/1.0.0/} \\
    \bottomrule
  \end{tabularx}
\end{table}

The atlas's text, figures and machine-readable file, and all results derived from VitalDB, are shared under
CC~BY~4.0, the licence of the data. Code is MIT. The atlas redistributes no recordings or per-case tables: results
are aggregates and figures, and a case is named only when it is the point of an example. It is a research
resource, not for clinical use, and is not affiliated with or endorsed by the VitalDB team, PhysioNet or any device
maker. Users should cite the atlas together with VitalDB and
PhysioNet~\citep{lee2022vitaldb,lee2022vitaldb-physionet,pollard2026physionet}.

There are three ways to use the atlas:
\begin{enumerate}[itemsep=2pt]
  \item \textbf{Read it.} Each variable page gives its measurand, delays, resolution and accuracy, and the chain of
    clocks behind its stamp. The synchronisation planner gives the expected misalignment between any two variables.
  \item \textbf{Load it.} The JSON file carries every parameter with its status and study, so an analysis can look
    up a delay or a drift rate, or propagate the stated distributions, rather than hard-code a number.
  \item \textbf{Apply it.} The corrected readers in vitaldb-bench return times on the working reference, with the
    labels attached.
\end{enumerate}

\section{How to contribute}
\label{sec:contribute}

Corrections, new studies, better models and code are welcome, and the atlas is built to be argued with. Each page
is one YAML file; a contribution is a pull request that edits or adds one, checked by
\texttt{python -m metrology\_atlas check}, which finds broken references, unknown fields, missing figures and
forbidden patterns such as unmarked case ids. Value models are versioned in vitaldb-resolution and never edited in
place: a refit becomes a new version, and the old one can still be called. When a result changes, the old study or
parameter is marked superseded rather than overwritten, so the history stays visible. Four rules apply everywhere:
name a case only when it is the point; publish only aggregates and figures; give every number a status and a
study; and supersede rather than overwrite.

The most useful contributions settle an open question. The atlas lists them in four kinds: phenomena seen but not
explained, explanations that the evidence only weakly supports, assumptions the release cannot test, and conflicts
where two studies disagree. Outside evidence would help most where the data cannot speak, above all the recorder
PCs' link to UTC and the hospital records' clock.

\section{Beyond VitalDB}
\label{sec:beyond}

The next atlases are other open physiological datasets, in particular multi-device waveform collections from
operating theatres and intensive care, where the same questions arise: how far apart the clocks are, what a
numeric's delay is, what a charted time or value really stands for. Much of the VitalDB work transfers directly:
the timing tools and the network adjustment are generic, and so is most of the value side. The rounding-model
engine, the distribution of a true value given a recorded one, and the model versioning in vitaldb-resolution do
not depend on VitalDB; only the registry of variables, a few recording rules and the fitted models do. We plan to
release that engine as a separate dataset-independent package, the value counterpart of physiosync, once a second
dataset shows which assumptions are general. Several findings (numerics that describe a window
before their stamp, rounded charted times, converters with fewer bits than documented) are likely to recur. A
shared schema also makes it possible to compare datasets on the same terms, for example how well two signals can
be aligned in each. Curators and researchers who would like to describe a dataset are invited to start from the
engine and the VitalDB folder as a template.

\section{Limitations and related work}

The atlas is a reconstruction. Every model is inferred from the records, without access to the devices, their
firmware or the recorder PCs, and none has been reviewed by the device makers. The chain to UTC is broken at its
first link, so absolute times are not traceable. Population results describe the release, not any one recording: a
per-setup or per-recording estimate is given where the data support one. Several results are preliminary and are
marked as such.

The work builds on the vocabulary and uncertainty framework of the GUM and the
VIM~\citep{jcgm-200-2012,jcgm-100-2008,jcgm-101-2008,jcgm-102-2011}, on statistical work on coarse and heaped
data~\citep{heitjan1989grouped,heitjan1991coarse,wang2008heaping}, and on network time
synchronisation~\citep{rfc5905}. It extends earlier work on timing errors in clinical
databases~\citep{goodwin2022timing}, EEG synchronisation in intensive care~\citep{goodwin2023}, data-quality
modelling of high-frequency physiological data~\citep{goodwin2023high}, pulse arrival time with bedside
equipment~\citep{ruffolo2025high}, and the storage of high-frequency signals~\citep{goodwin2020practical}. Delays in
processed EEG indices were measured by \citet{pilge2006timedelay}. To our knowledge, no open dataset has previously
been described as a complete measurement system in this way.

\section*{Acknowledgements}

We thank the VitalDB team for making their dataset openly available, and Tom Pollard (PhysioNet) for his advice
on releasing this work. Analysis code, figures and drafts of the text were
developed with the assistance of an AI system (Claude, Anthropic); the author designed the work, checked the
analyses and is responsible for all content.

\bibliographystyle{unsrtnat}
\bibliography{references}

\begin{thebibliography}{18}
\providecommand{\natexlab}[1]{#1}
\providecommand{\url}[1]{\texttt{#1}}
\expandafter\ifx\csname urlstyle\endcsname\relax
  \providecommand{\doi}[1]{doi: #1}\else
  \providecommand{\doi}{doi: \begingroup \urlstyle{rm}\Url}\fi

\bibitem[Goodwin et~al.(2022)Goodwin, Eytan, Dixon, Goodfellow, Doherty, Greer,
  McEwan, Tracy, Laussen, Assadi, and Mazwi]{goodwin2022timing}
A.~J. Goodwin, D.~Eytan, W.~Dixon, S.~D. Goodfellow, Z.~Doherty, R.~W. Greer,
  A.~McEwan, M.~Tracy, P.~C. Laussen, A.~Assadi, and M.~Mazwi.
\newblock Timing errors and temporal uncertainty in clinical databases -- a
  narrative review.
\newblock \emph{Frontiers in Digital Health}, 4:\penalty0 932599, 2022.
\newblock \doi{10.3389/fdgth.2022.932599}.

\bibitem[Goodwin et~al.(2023)Goodwin, Dixon, Mazwi, Hahn, Meir, Goodfellow,
  Kazazian, Greer, McEwan, Laussen, and Eytan]{goodwin2023}
A.~J. Goodwin, W.~Dixon, M.~Mazwi, C.~D. Hahn, T.~Meir, S.~D. Goodfellow,
  V.~Kazazian, R.~W. Greer, A.~McEwan, P.~C. Laussen, and D.~Eytan.
\newblock The truth {Hertz} -- synchronization of electroencephalogram signals
  with physiological waveforms recorded in an intensive care unit.
\newblock \emph{Physiological Measurement}, 44\penalty0 (8):\penalty0 085002,
  2023.
\newblock \doi{10.1088/1361-6579/ace49e}.

\bibitem[{Joint Committee for Guides in Metrology}(2012)]{jcgm-200-2012}
{Joint Committee for Guides in Metrology}.
\newblock International vocabulary of metrology -- {B}asic and general concepts
  and associated terms ({VIM}).
\newblock JCGM 200:2012, 2012.
\newblock \doi{10.59161/JCGM200-2012}.

\bibitem[{Joint Committee for Guides in
  Metrology}(2008{\natexlab{a}})]{jcgm-100-2008}
{Joint Committee for Guides in Metrology}.
\newblock Evaluation of measurement data -- {G}uide to the expression of
  uncertainty in measurement ({GUM}).
\newblock JCGM 100:2008, 2008{\natexlab{a}}.
\newblock \doi{10.59161/JCGM100-2008E}.

\bibitem[Lee et~al.(2022)Lee, Park, Yoon, Yang, Park, and Jung]{lee2022vitaldb}
H.-C. Lee, Y.~Park, S.~B. Yoon, S.~M. Yang, D.~Park, and C.-W. Jung.
\newblock {VitalDB}, a high-fidelity multi-parameter vital signs database in
  surgical patients.
\newblock \emph{Scientific Data}, 9:\penalty0 279, 2022.
\newblock \doi{10.1038/s41597-022-01411-5}.

\bibitem[Lee and Jung(2022)]{lee2022vitaldb-physionet}
H.-C. Lee and C.-W. Jung.
\newblock {VitalDB}, a high-fidelity multi-parameter vital signs database in
  surgical patients (version 1.0.0).
\newblock PhysioNet, 2022.
\newblock \doi{10.13026/czw8-9p62}.
\newblock URL \url{https://physionet.org/content/vitaldb/1.0.0/}.
\newblock RRID:SCR\_007345.

\bibitem[Pollard et~al.(2026)Pollard, Moody, Lehman, Gow, Fernandes, Xie,
  Johnson, Mark, and Heldt]{pollard2026physionet}
T.~Pollard, B.~E. Moody, L.~Lehman, B.~Gow, C.~Fernandes, C.~Xie, A.~Johnson,
  R.~G. Mark, and T.~Heldt.
\newblock {PhysioNet} as a global platform for biomedical research.
\newblock \emph{Nature Health}, 2026.
\newblock \doi{10.1038/s44360-026-00096-z}.

\bibitem[Lee and Jung(2018)]{lee2018vitalrecorder}
H.-C. Lee and C.-W. Jung.
\newblock Vital {Recorder} -- a free research tool for automatic recording of
  high-resolution time-synchronised physiological data from multiple
  anaesthesia devices.
\newblock \emph{Scientific Reports}, 8:\penalty0 1527, 2018.
\newblock \doi{10.1038/s41598-018-20062-4}.

\bibitem[{Joint Committee for Guides in Metrology}(2011)]{jcgm-102-2011}
{Joint Committee for Guides in Metrology}.
\newblock Evaluation of measurement data -- {S}upplement 2 to the ``{G}uide to
  the expression of uncertainty in measurement'' -- {E}xtension to any number
  of output quantities.
\newblock JCGM 102:2011, 2011.
\newblock \doi{10.59161/JCGM102-2011}.

\bibitem[{Joint Committee for Guides in
  Metrology}(2008{\natexlab{b}})]{jcgm-101-2008}
{Joint Committee for Guides in Metrology}.
\newblock Evaluation of measurement data -- {S}upplement 1 to the ``{G}uide to
  the expression of uncertainty in measurement'' -- {P}ropagation of
  distributions using a {Monte Carlo} method.
\newblock JCGM 101:2008, 2008{\natexlab{b}}.
\newblock \doi{10.59161/JCGM101-2008}.

\bibitem[Heitjan(1989)]{heitjan1989grouped}
D.~F. Heitjan.
\newblock Inference from grouped continuous data: a review.
\newblock \emph{Statistical Science}, 4\penalty0 (2):\penalty0 164--179, 1989.
\newblock \doi{10.1214/ss/1177012601}.

\bibitem[Heitjan and Rubin(1991)]{heitjan1991coarse}
D.~F. Heitjan and D.~B. Rubin.
\newblock Ignorability and coarse data.
\newblock \emph{The Annals of Statistics}, 19\penalty0 (4):\penalty0
  2244--2253, 1991.
\newblock \doi{10.1214/aos/1176348396}.

\bibitem[Wang and Heitjan(2008)]{wang2008heaping}
H.~Wang and D.~F. Heitjan.
\newblock Modeling heaping in self-reported cigarette counts.
\newblock \emph{Statistics in Medicine}, 27\penalty0 (19):\penalty0 3789--3804,
  2008.
\newblock \doi{10.1002/sim.3281}.

\bibitem[Mills et~al.(2010)Mills, Martin, Burbank, and Kasch]{rfc5905}
D.~Mills, J.~Martin, J.~Burbank, and W.~Kasch.
\newblock {Network Time Protocol} version 4: Protocol and algorithms
  specification.
\newblock RFC 5905, RFC Editor, 2010.
\newblock \doi{10.17487/RFC5905}.

\bibitem[Goodwin(2023)]{goodwin2023high}
A.~J. Goodwin.
\newblock \emph{High frequency physiological data quality modelling in the
  intensive care unit}.
\newblock PhD thesis, School of Biomedical Engineering, The University of
  Sydney, 2023.

\bibitem[Ruffolo et~al.(2025)Ruffolo, Siddiqui, Nguyen, Dixon, Assadi, Greer,
  Schwartz, Brudno, Mariakakis, and Goodwin]{ruffolo2025high}
I.~Ruffolo, A.~Siddiqui, B.~Nguyen, W.~Dixon, A.~Assadi, R.~Greer, S.~Schwartz,
  M.~Brudno, A.~Mariakakis, and A.~Goodwin.
\newblock High-fidelity measurement of pulse arrival time in critically ill
  children using standard bedside monitoring equipment.
\newblock \emph{Physiological Measurement}, 46\penalty0 (11):\penalty0 115006,
  2025.
\newblock \doi{10.1088/1361-6579/ae1b70}.

\bibitem[Goodwin et~al.(2020)Goodwin, Eytan, Greer, Mazwi, Thommandram,
  Goodfellow, Assadi, Jegatheeswaran, and Laussen]{goodwin2020practical}
A.~J. Goodwin, D.~Eytan, R.~W. Greer, M.~Mazwi, A.~Thommandram, S.~D.
  Goodfellow, A.~Assadi, A.~Jegatheeswaran, and P.~C. Laussen.
\newblock A practical approach to storage and retrieval of high-frequency
  physiological signals.
\newblock \emph{Physiological Measurement}, 41\penalty0 (3):\penalty0 035008,
  2020.
\newblock \doi{10.1088/1361-6579/ab7cb5}.

\bibitem[Pilge et~al.(2006)Pilge, Zanner, Schneider, Blum, Kreuzer, and
  Kochs]{pilge2006timedelay}
S.~Pilge, R.~Zanner, G.~Schneider, J.~Blum, M.~Kreuzer, and E.~F. Kochs.
\newblock Time delay of index calculation: analysis of cerebral state,
  bispectral, and {Narcotrend} indices.
\newblock \emph{Anesthesiology}, 104\penalty0 (3):\penalty0 488--494, 2006.
\newblock \doi{10.1097/00000542-200603000-00016}.

\end{thebibliography}

\end{document}